\documentclass[12pt,journal,compsoc]{IEEEtran}
\usepackage{cite}      

\usepackage{graphicx}  

\usepackage{psfrag}    

\usepackage{subfigure} 

\usepackage{url}       

\begin{document}
%
\title{Dynamics of Trends and Attention in Chinese Social Media}
%
%

\author{Louis Lei Yu\\
        Sitaram Asur\\
      Bernardo A. Huberman\\
\thanks{Louis Lei Yu, Department of Mathematics and Computer Science, Gustavus Adolphus College, 800 W College Ave, St Peter, MN 56082,  phone: (415)374-9197, FAX: (507) 933-7041, e-mail: lyu@gustavus.edu}
\thanks{Sitaram Asur, Social Computing Lab, HP Labs, 1501 Page Mill Road,  Palo Alto, CA 94304, phone: (650) 857-1501, fax: (650) 852-8156, email: sitaram.asur@hp.com}
\thanks{Bernardo A. Huberman, Social Computing Lab, HP Labs, 1501 Page Mill Road,  Palo Alto, CA 94304, phone: (650) 857-1501, fax: (650) 852-8156, email: bernardo.huberman@hp.com}}

\maketitle

\begin{abstract}
There has been a tremendous rise in the growth of online social networks all over the world in recent years.  It has facilitated users to generate a large amount of real-time content at an incessant rate, all competing with each other to attract enough attention and become popular trends. While Western online social networks such as Twitter have been well studied, the popular Chinese microblogging network Sina Weibo has had relatively lower exposure. In this paper, we analyze in detail the temporal aspect of trends and trend-setters in Sina Weibo, contrasting it with earlier observations in Twitter. We find that there is a vast difference in the content shared in China when compared to a global social network such as Twitter. In China, the trends are created almost entirely due to the retweets of media content such as jokes, images and videos, unlike Twitter  where it has been shown that the trends tend to have more to do with current global events and news stories.
We take a detailed look at the formation, persistence and decay of trends and examine the key topics that trend in Sina Weibo. One of our key findings is that retweets are much more common in Sina Weibo and contribute a lot to creating trends. When we look closer, we observe that most trends in Sina Weibo are due to the continuous retweets of a small percentage of fraudulent accounts. These fake accounts are set up to artificially inflate certain posts, causing them to shoot up into Sina Weibo's trending list, which are in turn displayed as the most popular topics to users.
\end{abstract}


\section{Introduction}

In the past few years, social media services as well as the users who subscribe to them, have grown at a phenomenal rate. 
This immense growth has been witnessed all over the world with millions of people of different backgrounds using these services on a daily basis. This widespread generation and consumption of content has created an extremely complex and competitive online environment where different types of content compete with each other for the attention of users. It is very interesting to study how certain types of content such as a viral video, a news article, or an illustrative picture,  manage to attract more attention than others, thus bubbling to the top in terms of popularity.  Through their visibility, these  popular topics contribute to the collective awareness reflecting what is considered important. It can also be powerful enough to affect the public agenda of the community.

There have been prior studies on the characteristics of trends and trend-setters in Western online social media (\cite{Asur2011}, \cite{Huberman}). In this paper, we examine in detail a significantly less-studied but equally fascinating online environment: Chinese social media, in particular, Sina Weibo: China's biggest microblogging network.

Over the years there have been news reports on various Internet phenomena in China, from the surfacing of certain viral videos  to the spreading of rumors (\cite{Jin}) to the so called ``human flesh search engines'':  a primarily Chinese Internet phenomenon of massive search using online media such as blogs and forums (\cite{flesh2}). These stories seem to suggest that many events happening in Chinese online social networks are unique products of China's culture and social environment.

Due to the vast global connectivity provided by social media, netizens all over the world are now connected to each other like never before; they can now share and exchange ideas with ease. It could be argued that  the manner in which the sharing occurs should be similar across countries.
However, China's unique cultural and social environment  suggests that the way individuals share ideas might be different than that in Western societies \cite{King}. For example, the age of Internet users in China is a lot younger. So it is likely that they may respond to different types of content than Internet users in Western societies. The number of Internet users in China is larger than that in the U.S, and the majority of users live in large urban cities. One would expect that the way these users share information can be even more chaotic. An important question to ask is to what extent would topics have to compete with each other in order to capture users' attention in this dynamic environment. Furthermore,  as documented by \cite{Tai}, it is known that the information shared between individuals in Chinese social media is monitored. Hence another interesting question to ask is what types of content would netizens respond to and what kind of popular topics would emerge u
 nder such constant surveillance. 

Given the above questions, we present an analysis on the evolution of trends in Sina Weibo. We monitored the evolution of the top trending keywords in Sina Weibo for 30 days. First, we analyzed the model of growth in these trends and examined the persistance of these topics over time. In this regard, we investigated if topics initially ranked higher tend to stay in the list of top 50 trending topics longer. 
Subsequently, by analyzing the timestamps of tweets, we looked at the propagation and decaying process of the trends in Sina Weibo and compare it to earlier observations of Twitter \cite{Asur2011}. 

Our findings are as follows:
\begin{itemize}
\item We discovered that the majority of trends in Sina Weibo are arising from frivolous content, such as jokes and funny images and photos unlike Twitter where the trends are mainly news-driven.
\item We established that retweets play a greater role in Sina Weibo than in Twitter, contributing more to the generation and persistence of trends. 
\item Upon examining the tweets in detail, we made an important discovery. We observed that many trending keywords in Sina Weibo are heavily manipulated and controlled by certain fraudulent accounts. The irregular activities by these accounts made certain tweets more visible to users in general.  
\item We found significant evidence suggesting that a large percentage of the  trends in Sina Weibo are  due to artificial inflation by fraudulent accounts.  The users we identified as fraudulent were 1.08\% of the total users sampled, but they were responsible for 49\% of the total retweets (32\% of the total tweets). 
\item We evaluated some methods to identify fraudulent accounts. After we removed the tweets associated with fraudulent accounts, the evolution of the tweets containing trending keywords follow the same persistent and decaying process as the one in Twitter. 
\end{itemize}

The rest of the paper is organized as follows. In Section \ref{background} we provide background information on  the development of Internet in China and on the Sina Weibo social network. In Section \ref{relate} we survey some related work on trends and spam in social media.  In Section \ref{evol}, we perform a detailed analysis of trending topics in Sina Weibo. In Section \ref{future}, we provide a discussion of our findings.

\section{Background } \label{background}

In this Section, we provide some background information on the Internet in China, the development of Chinese social media services, and Sina Weibo,  the most popular microblog service in China

\subsection{The Internet in China}

The development of the Internet industry in China over the past decade has been impressive. According to a survey from  the China Internet Network Information Center (CNNIC), by July 2008, the number of Internet users in China has reached 253 million, surpassing the U.S. as the world's  largest Internet market \cite{Statistic-general}. Furthermore, the number of Internet users in China as of 2010 was reported to be 420 million.

Despite this, the fractional Internet penetration rate in China is still low.  The 2010 survey by CNNIC on the Internet development in China \cite{Statistic-rural} reports that the Internet penetration rate in the rural areas of  China is on average $5.1\%$. In contrast,  the  Internet penetration rate in the urban cities of China is on average $21.6\%$. In metropolitan cities such as Beijing and Shanghai, the Internet penetration rate has reached over $45\%$, with Beijing being $46.4\%$ and  Shanghai  being $45.8\%$ \cite{Statistic-rural}. 

\begin{figure} [ht]
\centering
\includegraphics  [width=80mm, height=50mm]{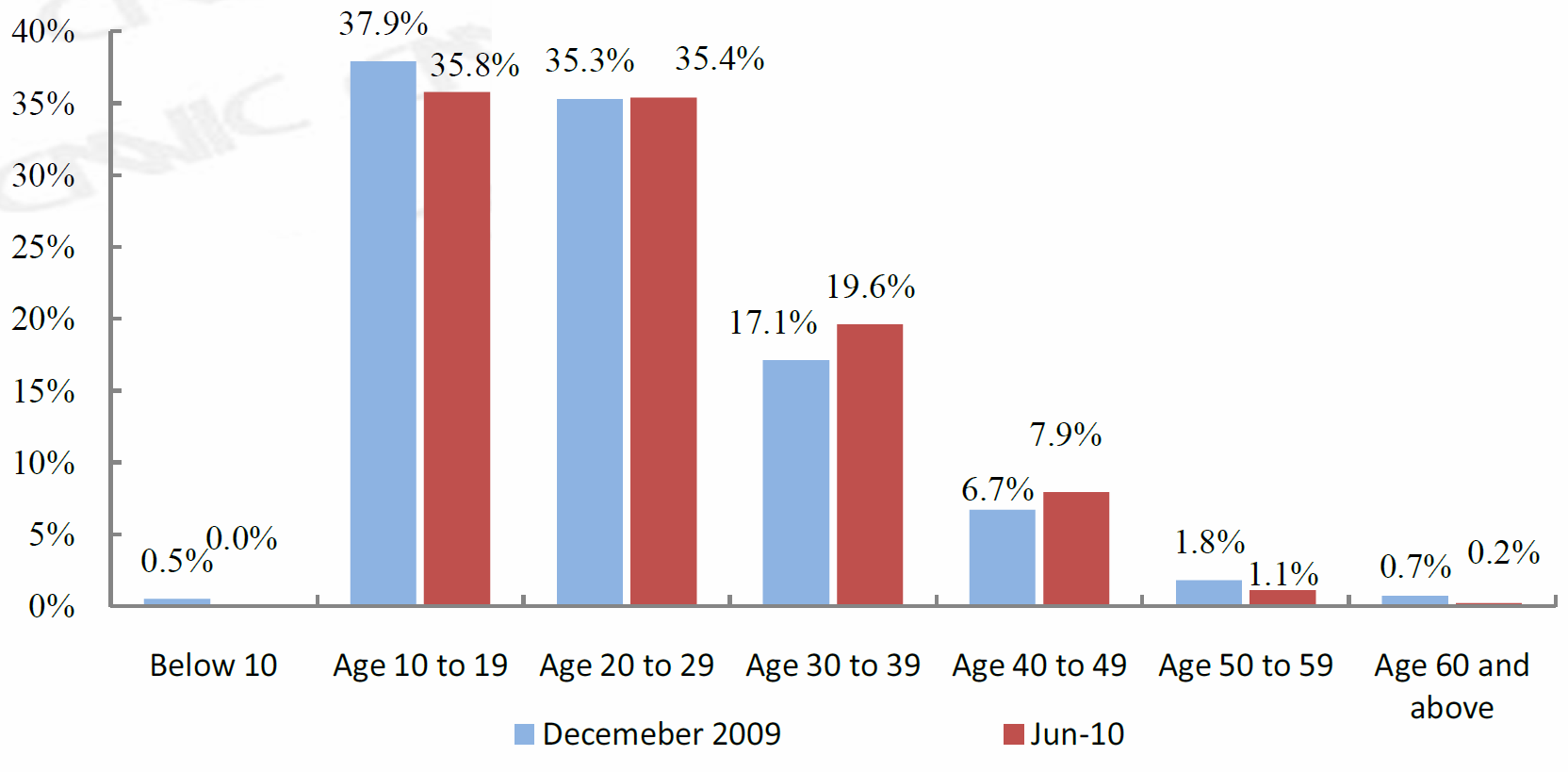}
\caption{ Age Distribution of Internet Users in China } \label{Age}
\end{figure}

\begin{figure} [ht]
\centering
\includegraphics [width=80mm, height=80mm]{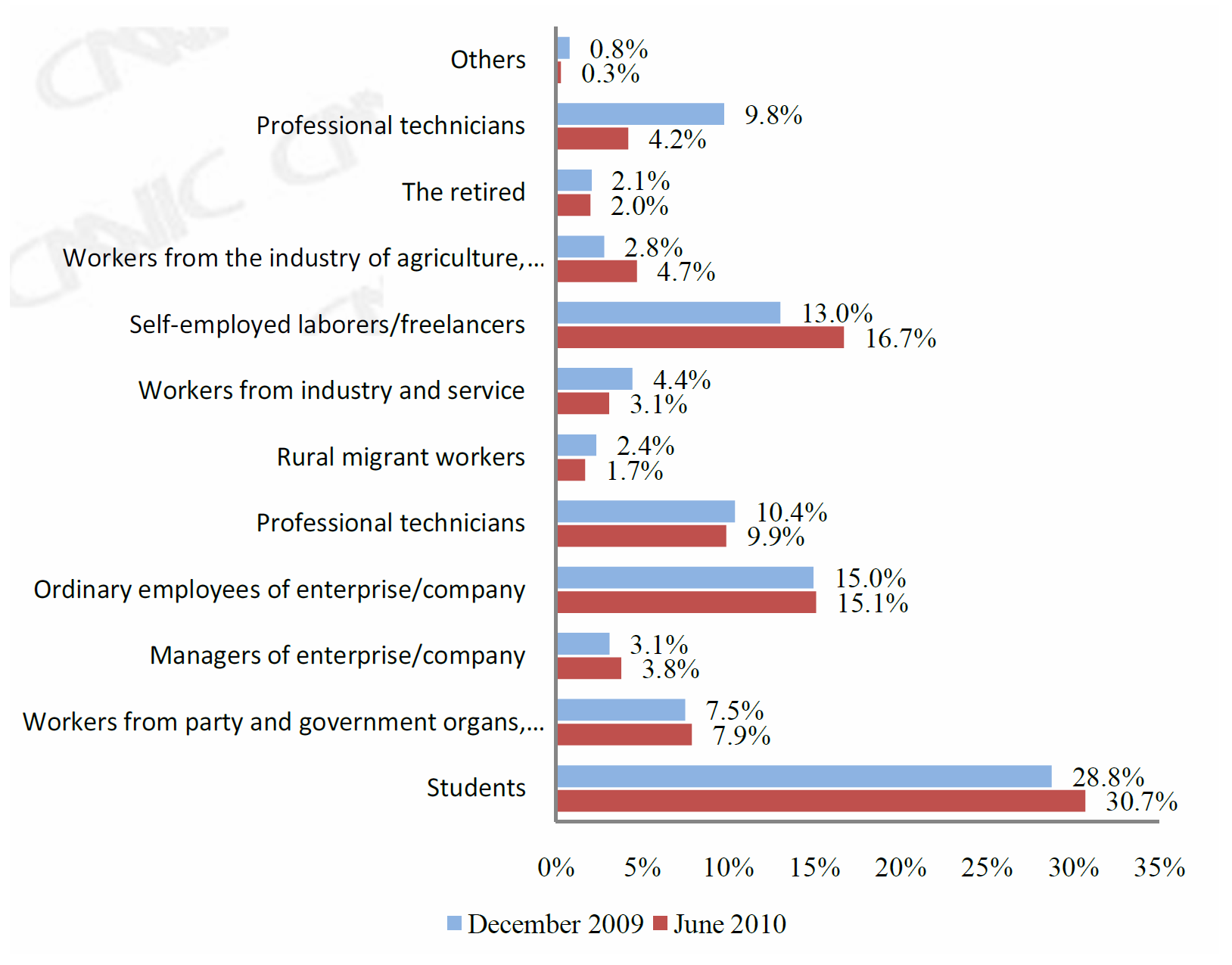}
\caption{ The Occupation Distribution of Internet Users in China } \label{Occupation}
\end{figure}

According to the survey by CNNIC in 2010 \cite{Statistic-general},  China's cyberspace is dominated by urban students between the age of 18--30 (see Figure \ref{Age} and Figure \ref{Occupation}, taken from \cite{Statistic-general}).  

The Government plays an important role in fostering the advance of the Internet industry in China.   Tai \cite{Tai} points out four major stages of Internet development in China, ``with each period reflecting a substantial change not only in technological progress and application, but also in the Government's approach to and apparent perception of the Internet.'' 

%
%
%
%
%

 According to  \textit{The Internet in China} \footnote{``The Internet in China'' by the Information Office of the State Council of the People's Republic of China is available at \textit{http://www.scio.gov.cn/zxbd/wz/201006/t667385.htm}}  released by  the Information Office of the State Council of China: 

 \begin{quote}  The Chinese government attaches great importance to protecting the safe flow of Internet information, actively guides people to manage websites in accordance with the law and use the Internet in a wholesome and correct way. \end{quote}

\subsection{Chinese Online Social Networks}

Online social networks are a major part of the Chinese Internet culture \cite{Jin}.  Netizens\footnote{A netizen is a person actively involved in online communities \cite{Netizen}.}  in China organize themselves using  forums, discussion groups, blogs, and social networking platforms to engage in  activities such as exchanging viewpoints and sharing information \cite{Jin}.  According to \textit{The Internet in China}:
 
 \begin{quote} 
 
 Vigorous online ideas exchange is a major characteristic of China's Internet development, and the huge quantity of BBS posts and blog articles is far beyond that of any other country. China's websites attach great importance to providing netizens with opinion expression services, with over 80\% of them providing electronic bulletin service. In China, there are over a million BBSs and some 220 million bloggers. According to a sample survey, each day people post over three million messages via BBS, news commentary sites, blogs, etc., and over 66\% of Chinese netizens frequently place postings to discuss various topics, and to fully express their opinions and represent their interests. The new applications and services on the Internet have provided a broader scope for people to express their opinions. The newly emerging online services, including blog, microblog, video sharing and social networking websites are developing rapidly in China and provide greater convenience for Chi
 nese citizens to communicate online. Actively participating in online information communication and content creation, netizens have greatly enriched Internet information and content.
  \end{quote}

\subsection{Sina Weibo}
Sina Weibo was  launched by the Sina corporation, China's biggest web portal, in August 2009.   It has been reported by the Sina corporation  that Sina Weibo now has 250 million registered accounts and generates 90 million posts per day. Similar to Twitter, a user profile in Sina Weibo displays the user's name, a brief description of the user, the number of followers and followees the user has. 
There are three types of user accounts in Sina Weibo, regular user accounts, verified user accounts, and the expert (star) user account.  A verified user account typically represents a well known public figure or organization in China.   


Twitter users can address tweets to other users and can mention others in their tweets.  A  common practice in Twitter is ``retweeting'',  or rebroadcasting someone else's messages to one's followers.  The equivalent of a retweet in Sina Weibo is instead shown as two amalgamated entries: the original entry and the current user's actual entry which is a commentary on the original entry. 

Sina Weibo  has another functionality absent from Twitter: the comment. When a Sina Weibo user makes a comment, it is not rebroadcasted to the user's followers. Instead, it can only be accessed under the original message.

\section{Related Work} \label{relate}

In this Section, we provide a survey of papers in two related areas: spam detection and the study of trends in social networks. In each area, we present work on both Western social networks and Chinese social networks. 

\subsection{Spam Detection in Twitter}

Spam and bot detection in social networks is a relatively recent area of research, motivated by the vast popularity of social websites such as Twitter and Facebook.  It draws on research from several areas of computer science such as computer security, machine learning, and network analysis.

In the 2010 work by Benevenuto et al \cite{benevenuto2010detecting}, the authors examine spam detection in Twitter by first collecting a large dataset of more than 54 million users, 1.9 billion links, and 1.8 billion tweets. After exploring content and behavoir attributes, they developed an SVM classifier and was able to detect spammers with 70\% precision and non-spammers with 96\% precision. As an insightful follow up, the authors used $\chi^2$ statistics to evaluate the importance of the attributes they used in their model. 

The second paper with direct application to spam detection in Twitter was by Wang \cite{wang2010detecting}. Wang motivated his research with the statistic that an estimated 3\% of messages in Twitter are spam. The dataset used in in this study was relatively smaller, gathering information from 25,847 users, 500 thousand tweets, and 49 million follower/friend relationships. Wang used decision trees, neural network, SVM, and naive Bayesian models.

Finally, Lee et al. \cite{lee2010uncovering} described a different approach to detect spammers. They created honeypot user accounts in Twitter and recorded the features of users who interact with these accounts. They then used these features to develop a classifier with high precision. 

\subsection{Spam Detection in General Online Social Networks}

In social bookmarking websites,  Markines et al. \cite{markines2009social} used just  6 features - tag spam, tag blur, document structure, number of ads, plagiarism, and valid links,  to develop a classifier with 98\% accuracy.

On facebook,  Boshmaf et al.  successfully launched a network of social bots \cite{bgh1982}. Despite Facebook's bot detection system, the authors were able to achieve an 80\% infiltration rate over 8 weeks. 

In online ad exchanges, advertisers pay websites for each user that clicks through an ad to their website. The way fraud occurs in this domain is for bots to click through ads on a website owned by the botnet owners. The money at stake in this case has made the bots employed very sophisticated. The botnet owners use increasingly stealthy, distributed traffic to avoid detection. Stone et al. examined various attacks and prevention techniques in cost per click ad exchanges \cite{stone2011understanding}.  Yu et al. \cite{yu2010sbotminer} gave a sophisticated approach to detect low-rate bot traffic by developing a model that examines query logs to detect coordination across bots within a botnet.

\subsection{Spam Detection in Chinese Online Social Networks}

Some studies had been done on  spam and bot detection in Chinese online social networks \cite{6406086},  \cite{6425674}.   Xu et al. \cite{XuChen} observed the spammers in Sina Weibo and found that the spammers can be classified into two categories: promoters and robot accounts. 

Lin et al. \cite{Lin:2013:AIS:2501025.2501035} presented an analysis of spamming behaviors in Sina Weibo. Using methods such as proactive honeypots, keyword based search and buying spammer samples directly from online merchants. they were able to collect a large set of spammer samples. Through their analysis they found three representative spamming behaviors: aggressive advertising, repeated duplicate reposting, and aggressive following. 


  spammer identification system. Through tests with real data it is demonstrated that the system can effectively detect the spamming behaviors and identify spammers in Sina Weibo.

\subsection{Battling the ``Internet Water Army'' in Chinese Online Social Networks}

One relevant area of research is the study of the ``Online Water Army'' \footnote {e.g., \textit{http://shuijunwang.com} or \textit{http://www.51shuijun.net}}.	It represents full-time or part-time paid posters hired by PR companies to help in raising the popularity of a specific company or person by posting articles, replies, and comments in online social networks.  According to CCTV  \footnote {see report in Chinese at \textit{http://news.cntv.cn/china/20101107/102619.shtml}}, these paid posters in China help their customers using one of the following three tactics: 1. promoting a specific product, company or  person; 2. smear/slander competitors;  3. help deleting negative posts or comments.


 st in BBS systems, and online social networks.

In the work by Chen et al. \cite{ChenWu}, the authors examined comments in the Chinese news websites such as Sina.com and Sohu.com and used reply, activity, and semantic features to develop an SVM classifier via the LIBSVM Python library with 95\% accuracy at detecting paid posters. Interesting information discussed in the paper includes the organizational structure of PR firms which hire  the paid posters and the choice of features: percentage of replies, average interval time of posts, active days, and number of reports commented on.

\subsection{Measuring Influences  in Online Social Networks}

For many years the structural properties of various Western social networks have been well studied by sociologists and computer scientists \cite{Jamali} \cite{Mislove} \cite{Buchanan} \cite{Kumar}.

In social network analysis, \textit{social influence} refers to the concept of people modifying their behavior to bring them closer to the behavior of their friends. In a social-affiliation network consists of nodes representing individuals,  links representing friendships, and nodes representing \textit{foci}: ``social, psychological, legal, or physical entities around which joint activities are organized (e.g., workplace, social groups) \cite{mcpherson2001birds}'',  if $A$ and $B$ are friends, and $F$ is a focus that $A$ participates in.  Over time, $B$ can participate in the same focus due to $A$'s involvement, this is called a \textit{membership closure}\cite{mcpherson2001birds}.

Agarwal et al. ~\cite{Agarwal2008Identifying}  examined methods to identify influential bloggers in the blogosphere. They discovered that the most influential bloggers are not necessarily the most active.  Backstrom et al. \cite{Backstrom} studied the characteristics of \textit{membership closure} in LiveJournal. Crandall et al. \cite{Crandall} studied the adaptation of influences between editors of Wikipedia articles. 

Romero et al. ~\cite{Romero2011} measured retweets in Twitter and found that passivity was a major factor when it comes to message forwarding. Based on this result, they  presented a measure of social influences that takes into account the passivity of the audience in social networks.

\subsection{The Study of Trends in Twitter}
There are various studies on trends in Twitter \cite{Huberman} \cite{Kwak} \cite{Mathioudakis} \cite{Wu2}.

One of the most extensive investigations into trending topics in Twitter was by Asur et al. \cite{asur2011trends}. The authors examined the growth and persistence of trending topics in Twitter and observed that it follows a log-normal distribution of popularity. Accordingly, most topics faded from popularity relatively quickly, while a few topics lasted for long periods of time. They estimated the average duration of topics to be around 20-40 minutes.
When they examined the content of the trends, they observed that traditional notions of influence such as the frequency of posting and the number of followers were not the main drivers of popularity in trends. Rather it was the resonating nature of the content that was important. An interesting finding was that news topics from traditional media sources such as CNN, New York Times and ESPN was shown to be some of the most popular and long lasting trending topics in Twitter, suggesting that Twitter amplifies some of the broader trends occurring in society.

Cha et al.  \cite{cha2010measuring} explored user influences on Twitter trends and discovered some interesting results. First, users with many followers were found to not be very effective in generating mentions or retweets. Second, the most influential users tend to influence more than one topic. Third, influences were found to not arise spontaneously, but instead as the result of focused efforts, often concentrating on one topic.

%
%
%
%

\subsection{Social Influences and the Propagation of Information in  Chinese Social Networks}

Researchers have analyzed the structure of various Chinese offline social networks \cite{StrongTie} \cite{WorkControl} \cite{768262} \cite{Guanxi3} \cite{Carrington}.  

There have been only a few studies on social influences  in Chinese online social networks. Jin \cite{Jin}  studied the structure and interface of Chinese online Bulletin Board Systems (BBS) and the  behavioral patterns of its users.  Xin \cite{Xin} conducted a survey of BBS's influence on University students in China.  Yu et al.  \cite{yu}  looked at the adaptation of books, movies, music, events and discussion groups on Douban, the largest online media database and one of the largest online communities in China.

In a similar area, there are some studies on the structural properties and the characteristics of information propagation in Chinese online social networks \cite{Zhong2010}  \cite{Zhang201215}, \cite{Chan2012}, \cite{Chen2012}, \cite{Chu2011}. Yang et al. \cite{Yang} noted that various information services  (e.g., eBay, Orkut, and Yahoo!)  encountered serious challenges when entering China. They presented an empirical study of social interactions among Chinese netizens based on over 4 years of comprehensive data collected from Mitbbs (www.mitbbs.com),  the most frequently used online forum for Chinese nationals who are studying or working abroad.
 
 Lin et al. \cite{Lin} presented a comparison of  the interaction patterns between two of the largest online social networks in China: Renren and Sina Weibo.  Niu et al. \cite{Renren} gave an empirical analysis of Renren, it follows an exponentially truncated power law in-degree distribution, and has a short average node distance.

 King et al. \cite{King} studied the concept of \textit{guanxi}, a unique dyadic social construct, as applied to the interaction between web sites in China.  Chang et al. \cite{Chang2} studied a special case of the propagation of information in Chinese online social networks: the sending and receiving of messages containing wishes and moral support. They  provided analysis on the data from Linkwish, a micro social network for wish sharing with users mainly from Taiwan, Hong Kong, and Macao. 
 
 Fan et al. \cite{Fan} looked at the propagation of emotion in Sina Weibo.  They found that the correlation of anger among users is significantly higher than that of joy, which indicates that angry emotion could spread more quickly and broadly in the network. And, there is a stronger sentiment correlation between a pair of users if they share more interactions. Finally,  users with larger number of friends possess more significant sentiment influence to their neighborhoods.


\section{Analysis of Trends and Trend-Setters in Sina Weibo} \label{evol}
\subsection{The Trending Keywords}

Sina Weibo offers a list of 50 keywords that appear most frequently in users' tweets. They are ranked according to the frequency of appearances in the last hour. This is similar to Twitter, which also presents a constantly updated list of trending topics: keywords that are most frequently used in tweets over a period of time. We extracted these keywords over a period of 30 days  (from June 18th, 2011 to July 18th, 2011) and retrieved all the corresponding tweets containing these keywords from Sina Weibo.

\begin{figure*} [ht]
\centering
\includegraphics[width=160mm, height=80mm] {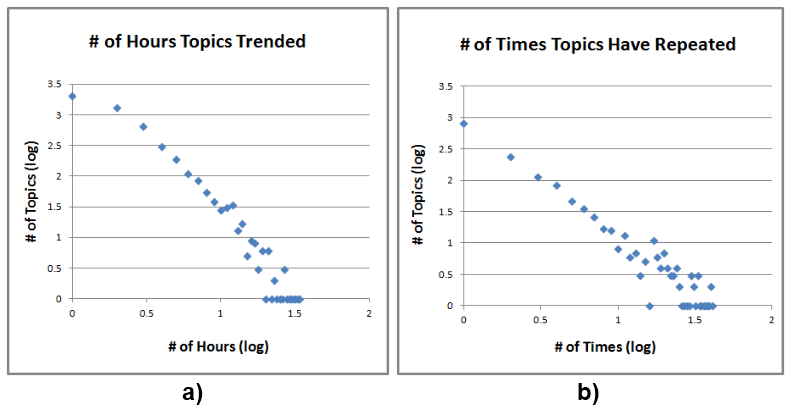}
\caption{ Distributions of trending time and the number of times topics reappeared } \label{power}
\end{figure*}

We first monitored the hourly evolution of the top 50 keywords in the trending list for 30 days. We observed that the average time spent by each keyword in the hourly trending list is 6 hours. And the distribution for the number of hours each topic remains on the top 50 trending list follows the power law (as shown  in Figure \ref{power} a). The distribution suggests that only a few topics exhibit long-term popularity. 
Another interesting observation is that a lot of the key words tend to disappear from the top 50 trending list after a certain amount of time and then later reappear. We examined the distribution for the number of times  keywords reappear in the top 50 trending list (Figure \ref{power} b). We observe that this distribution follows the power law as well.

Both the above observations are very similar to the earlier study of trending topics in Twitter by \cite{Asur2011}. However, one important difference with Twitter is that the average trending time is significantly higher in Sina Weibo (in Twitter it was 20-40 minutes). This suggests that Weibo may not have as many topics competing for attention as Twitter.

\begin{figure} [ht]
\centering
\includegraphics[width=80mm, height=60mm] {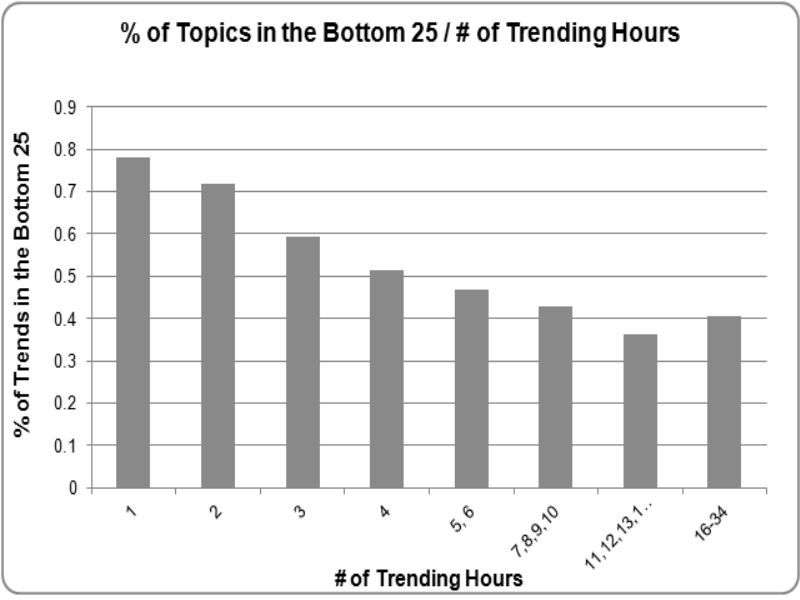}
\caption{Distribution of trending times for topics in the bottom 25 of the top 50 trend list} \label{stay}
\end{figure}

Following our observation that some keywords stay in the top 50 trending list longer than others, we wanted to investigate if topics that are ranked higher initially tend to stay in the top 50 trending list longer.  We separated the top 50 trending keywords into two ranked sets of 25 each: the top 25 and the bottom 25. Figure \ref{stay} illustrates the plot for the percentage of topics that placed in the bottom 25 relating to the number of hours these topics stayed in the top 50 trending list. We can observe that topics that do not last are usually the ones that are in the bottom 25. On the other hand, the long-trending topics spend most of their time in the top 25, which suggests that items that become very popular are more likely to stay longer in the top 50. This intuitively means that items that attract phenomenal attention initially are not likely to dissipate quickly from people's interests.

\subsection{The Evolution of Tweets} 
 
Next, we investigate the process of persistence and decay for the trending topics in Sina Weibo. In particular, we want to measure the distribution for the time intervals between tweets containing the trending keywords. We continuously monitored the keywords in the top 50 trending list and for each trending topic we retrieved all the tweets containing the keyword from the time the topic first appeared in the top 50 trending list until the time it disappeared. Accordingly, we collected complete data for 811 topics over the course of 30 days (from June 20th, 2011 to July 20nd, 2011). 
In total we collected 574,382 tweets from 463,231 users.  Among the 574,382 Tweets,  35\% of the tweets (202,267 tweets) are original tweets, and 65\% of the tweets (372,115 tweets) are retweets. 40.3\% of the total users (187130 users)  retweeted at least once in our sample. 

We measured the number of tweets that each topic gets in 10 minute intervals, from the time the topic starts trending until the time it stops. From this we can sum up the tweet counts over time to obtain the cumulative number of tweets $N_q(t_i)$ of topic $q$ for any time frame $t_i$,
This is given as :
\begin{equation}
\large {N_q(t_i) = \sum_{\tau = 1}^{i} n_q(t_\tau)}
\end{equation}

where $n_q(t)$ is the number of tweets on topic $q$ in time interval $t$. We then calculate the ratios
$C_q(t_i, t_j) = N_q(t_i) / N_q(t_j)$ for topic $q$ for time frames $t_i$ and
$t_j$.  

\begin{figure*} [ht]
\centering
\includegraphics[width=160mm, height=80mm] {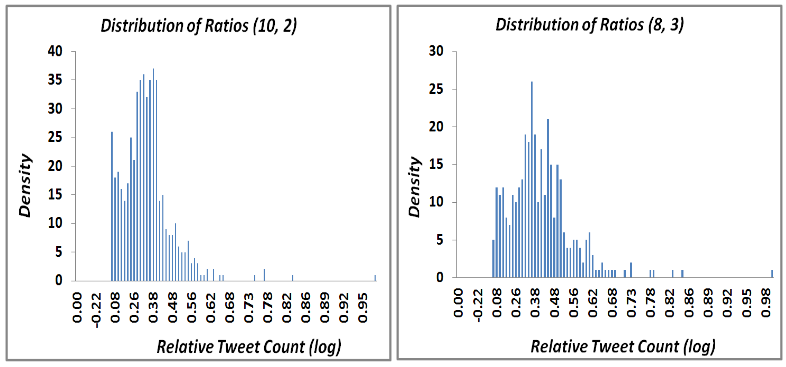}
\caption{The distribution of $C_q(t_i, t_j)$'s over all topics for two arbitrarily chosen pairs of time frames: (10, 2) and (8, 3)} \label{ratio_general}
\end{figure*}

Figure \ref{ratio_general} shows the distribution of $C_q(t_i, t_j)$'s over all topics for two arbitrarily chosen pairs of time frames: (10, 2) and (8, 3) (nevertheless such that $t_i > t_j$, and $t_i$ is relatively large, and $t_j$ is small).

These figures suggest that the ratios $C_q(t_i, t_j)$ are distributed according to the log-normal distributions. We tested and confirmed that the distributions indeed follow the log-normal distributions.  

This finding agrees with the result from a similar experiment in Twitter trends. Asur and others \cite{Asur2011} argued that the log-normal distribution occurs due to the multiplicative process involved in the growth of trends which incorporates the decay of novelty as well as the rate of propagation. The intuitive explanation is that at each time step the number of new tweets (original tweets or retweets) on a topic is multiplied over the tweets that we already have. The number of past tweets, in turn, is a proxy for the number of users that are aware of the topic up to that point. These  users  discuss the topic on different forums, including Twitter, essentially creating an effective network through which the topic spreads. As more users talk about a particular topic, many others are likely to learn about it, thus giving the multiplicative nature of the spreading.  On the other hand, the monotically decreasing decaying process characterizes the decay in timeliness and nove
 lty of the topic as it slowly becomes obsolete.

However, while only  35\% of the tweets in Twitter are retweets, there is a much larger percentage of tweets that are retweets in Sina Weibo. From our sample we observed that a high 65\% of the tweets are retweets. This implies that the topics are trending mainly because of some content that has been retweeted many times. Thus, Sina Weibo users are more likely to learn about a particular topic through retweets. 

\subsection{Trend-setters in Sina Weibo}

For every new trending keyword we retrieved the most retweeted tweets in the past hour and compiled a list of most retweeted users. 
Table \ref{Trend_Setter_Retweet} illustrates the top 20 most retweeted authors appearing in at least 10 trending topics each. The influential authors are ranked according to the ratio between the number of times the authors' tweets are retweeted and the number of trending topics these tweets appeared in.  


\begin{table*}[ht]
\caption{Top 20 Retweeted Users in At Least 10 Trending Topics} \label{Trend_Setter_Retweet}
\centering 
\begin{tabular}{|c|c|c|c|c|c|}
	\hline
	& Account Descriptions    & Verified  &  \# of Times Retweeted  & \# of Tweets &  \# of Topics \\
	\hline
 1& Fashion Magazine & Yes & 1194999 &  37 & 12  \\
2&   Fashion Brand  &   Yes &   849404 & 21  & 13   \\
3&Travel Magazine &  Yes & 127737 & 123 & 21 \\
4& Gourmet Factory & No & 553586 & 86 & 12 \\
5& Horoscopes & No & 1545955 & 101 & 38\\
 6& Silly Jokes& No & 3210130 & 258 & 81\\
 7&  Good Movies & No & 1497968	& 140 & 38  \\
8&  Wonderful Quotes	& No &	602528	&39& 17\\
9&	Global Music &	No&	697308	&116	&22	\\
10&	Funny Jokes  &No&	3667566 &	438&	121	\\
11	& Creative Ideas 	& No & 	742178	& 111	& 25	\\
12	& Chinese singer & Yes & 	284600	& 25	 & 10 \\
13	& Good Music	&No&	323022	& 52	& 12	\\
14	& Movie Factory	& No & 	1509003	&230	&59	\\
15	&Strange Stories	&No &	1668910	&250	&66	\\
16	&Beautiful Pictures		&No&	435312&	33&	18\\
17	&Global Music		&No	&432444	&65	&18	\\
18	&Female Fashion		&No	&809440	&87	&34	\\
19	&Useful Tips &	No&	735070&	153&	31\\
20	&Funny Quizzes 		&No	&589477&	77&	25\\
	\hline	
\end{tabular}
\end{table*}

From Table  \ref{Trend_Setter_Retweet} we observed that only 4 out of the top 20 influential authors were verified accounts. The 4 verified accounts represent an urban fashion magazine, a fashion brand, an online travel magazine, and a Chinese celebrity. The  other 16 influential authors are unverified accounts. They all seem to have a strong focus on collecting user-contributed jokes, movie trivia, quizzes, stories and so on. This is in sharp contrast to 
the topics that are popular in Twitter as reported by \cite{Asur2011}. When we looked at a longer list of authors we observed the same trend. The most popular items were all related to frivolous content and media, unlike Twitter which had a strong affinity towards news and current events.

The ``\# of Times Tweeted'' column in Table~\ref{Trend_Setter_Retweet} gives the unique tweets that have been retweeted. We can observe that the rate at which they have been retweeted is phenomenal. For example, the top retweeted user posted 37 tweets which in total were retweeted 1194999 times.

\subsection{The Evolution of Retweets and Original Tweets}
\begin{figure*} [ht]
\centering
\includegraphics[width=160mm, height=140mm] {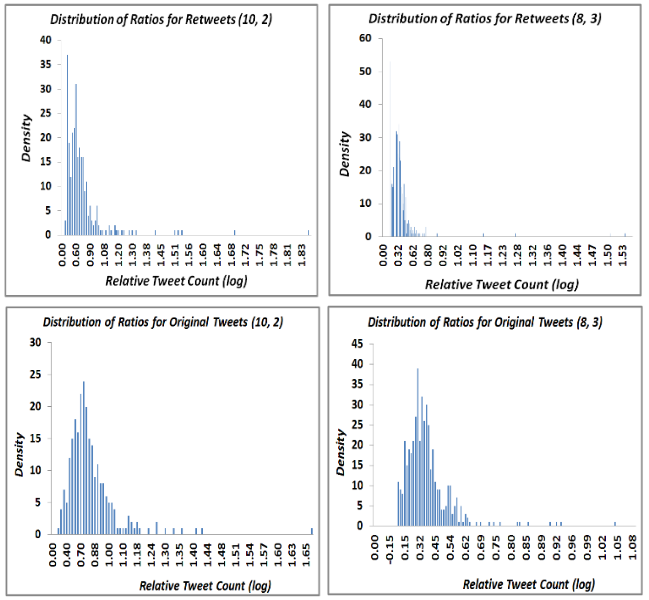}
\caption{The densities of ratios between cumulative original tweets/retweets counts measured in two arbitrary time frames:  (10, 2) and (8, 3)} \label{ratio_single_retweet}
\end{figure*}

In the next experiemnt, we separate the tweets in Sina Weibo into original tweets and retweets and calculate  the densities of ratios between cumulative retweets/original tweets counts measured in different time frames. Figure \ref{ratio_single_retweet} shows the distributions of original tweets/retweets ratios over all topics for two arbitrarily chosen pairs of time frames: (10, 2) and (8, 3).

We find (as the last two sub-figures in Figure \ref{ratio_single_retweet} show) that the distributions of ratios for original tweets follow the log-normal distribution. However, we observe (as the first two sub-figures in Figure  \ref{ratio_single_retweet} show) that for retweets, the distributions do not satisfy all the properties of the log-normal distribution. This is indicated by the large amount of low retweet ratios in the distribution.  Furthermore, there are high spikes in the lower ratios area of the distribution. 

\subsection{ Identifying Spam Activity in Sina Weibo}

From Figure \ref{ratio_single_retweet} in the previous Section we observed that there is a high percentage of low ratios in the distribution of retweet ratios. This suggests that for a lot of the topics, there is an initial flurry of retweets. We hypothesize that this is due to the activities of certain users in Sina Weibo. As these accounts post a tweet, they tend to set up many other fake accounts to continuously retweet this tweet, expecting that the high retweet numbers would propel the tweet to place in the Sina Weibo hourly trending list. This would then cause other users to notice the tweet more after it has emerged as the top hourly, daily, or weekly trend setter.  We attempt to verify the above hypothesis empirically.  We define a {\bf spamming account} as one that is set up for the purpose of repeatedly retweeting certain messages, thus giving these messages artificially inflated popularity. According to our hypothesis,  the users who retweet abnormally high amounts
   are more likely to be spam accounts. 

\begin{figure*} [ht]
\centering
\includegraphics[width=160mm, height=80mm] {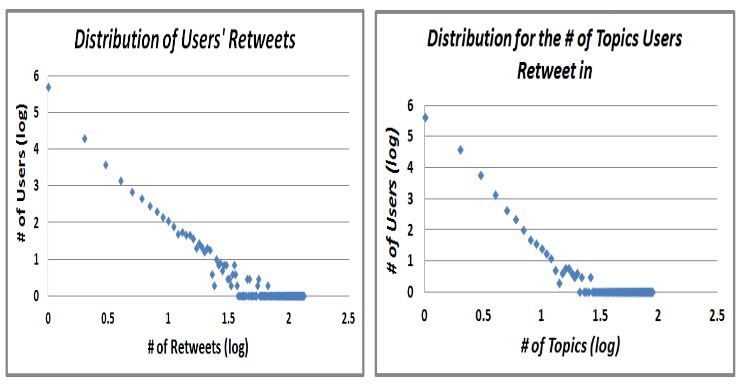}
\caption{The distribution for the number of users' retweets and the number of topics users' retweets trend in } \label{distribution_r_r}
\end{figure*}

\begin{figure} [ht]
\centering
\includegraphics[width=80mm, height=60mm] {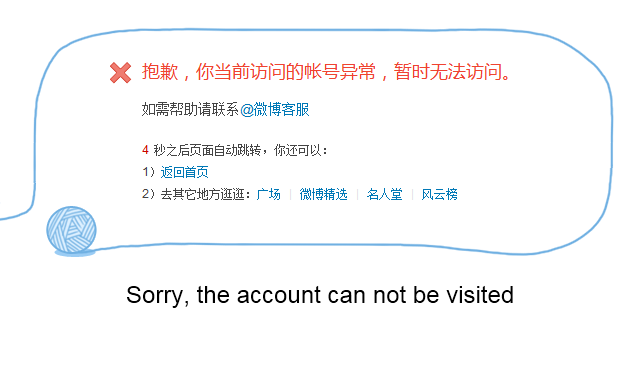}
\caption{An Example of an Error Page} \label{error}
\end{figure}

Figure \ref{distribution_r_r} a) illustrates the distribution for the number of users and their corresponding number of retweets (over all topics). Figure \ref{distribution_r_r} b) illustrates the distribution for the number of users and the numbers of topics that they caused to trend by their retweets. We observe that both distributions in Figure \ref{distribution_r_r}  follow the power law. This implies that there are certain users who retweet a lot, and a small number of users are responsible for a large number of topics. Next, we investigate who these users are.
We manually checked the top 40 accounts who retweeted the most. To our surprise, 37 of these 40 accounts could no longer be accessed. That is, when we queried the accounts' IDs, we retrieved a message from Sina Weibo stating that the account has been removed and can no longer be accessed (see Figure \ref{error}). 

According to Sina Weibo's frequently asked question page, if a user sends a tweet containing illegal or sensitive information, such tweet will be immediately deleted by Sina Weibo's administrators, however, the users' accounts will still be active. For the above reason we assume that if an account was active one month ago and can no longer be reached, it indicates that this account has very likely performed malicious activities such as spamming and has hence been deleted.

Next, we inspect the user accounts with the most retweets in our sample and the number of accounts they retweeted.  We see that although these accounts retweeted a lot, they mostly only retweet messages from a few users. We re-organize the users who retweeted by the ratio between the number of times he/she retweeted and the number of users he/she retweeted. We refer to this as the {\bf user-retweet ratio}. Table \ref{verified} illustrates the top 10 users with the highest user-retweet ratios.  We note that for all these users, they  each retweet posts from only one account. We observe that this is true for the top 30 accounts with the highest user-retweet ratios. 

\begin{table*}[ht]
\caption{The top 10 accounts with the highest user-retweet ratios (u-r ratio)}
\centering
\begin{tabular}{|c|c|c|c|c|}
\hline
User ID & Number of  Retweets & Number of Users Retweeted & U-R Ratio \\
\hline
1840241580	&134		&1&134\\
2241506824	&125	&1&125\\		
1840263604	&68		&1&68	\\		
1840237192	&64	&1&64		\\		
1840251632	&64		&1&64	\\		
2208320854	&55		&1& 55	\\		
2208320990	&51	&1&51		\\		
2208329370	&48		&1&48	\\		
2218142513	&47		&1&47	\\		
1843422117	&44		&1&44	\\			
\hline
\end{tabular}
\label{verified}
\end{table*}

Next, we conduct the following experiment: starting from the users with the highest user-retweet ratios, we used a crawler to automatically visit and retrieve each user's Sina Weibo account. Thus we measured the percentage of user accounts that can still be accessed (as opposed to be directed to the error page) organized by user-retweet ratios  (Table \ref{still}).  We observe that only 12\% of the accounts with user-retweet ratios of above 30 are active. And, 
as user-retweet ratios decrease, the percentages of active accounts slowly increase. We consider this to be strong evidence for the hypothesis that user accounts with high user-retweet ratios are likely to be spam accounts.  

\begin{table*}[ht]
\caption{The percentage of accounts whose profiles can still be accessed, organized by user-retweet ratio}
\centering
\begin{tabular}{|c|c|c|}
\hline
Ratio&	Percentage of Active Accounts&		Percentage of Inactive Accounts\\	
\hline		
$\geq$30	&12\%&	88\%\\			
20 -- 29	&38\%	&63\%\\			
11 -- 19&	16\%&	84\%\\			
10&	22\%&		78\%\\			
9&	12\%&	88\%\\			
8&	16\%&	84\%\\			
7&	15\%&	85\%\\			
6&	21\%&	79\%\\			
5&	30\%&	70\%\\			
4&	58\%&		42\%\\			
3&	80\%&	20\%\\			
2&	96\%&	4\%\\
1& 92\% & 8\%\\		
\hline
\end{tabular}
\label{still}
\end{table*}

We observe that in some cases, accounts with lower user-retweet ratios can still be a spam account. For example, an account could retweet a number of posts from other spam accounts, thus minimizing the suspicion of being detected as a spam account itself.  

\subsection{Removing Spammers in Sina Weibo}

From our sample, after automatically checking each account,  we identified 4985 accounts that were deleted by the Sina Weibo administrator.  We called these 4985 accounts ``suspected spam accounts''. There were 463,231 users in our sample, and 187,130 of them retweeted at least once. Thus we identified 1.08\% of the total users (2.66\% of users that retweeted) as suspected spam accounts. 

Next, in order to measure the effect of spam on the Weibo network, we removed all retweets from our sample disseminated by suspected spam accounts as well as posts published by them (and then later retweeted by others).  We hypothesize that by removing these retweets, we can eliminate the influences caused by the suspected spam accounts. We observed that after these posts were removed, we were left with only 189,686 retweets in our sample (51\% of the original total retweets). In other words, by removing retweets associated wth suspected spam accounts, we {\bf successfully removed 182,429 retweets, which is 49\% of the total retweets and 32\% of total tweets (both retweets and original tweets) from our sample}.  This result is very interesting because it shows that a large amount of retweets in our sample are associated with suspected spam accounts. The spam accounts are therefore artificially inflating the popularity of topics, causing them to trend. 

\begin{figure} [ht]
\centering
\includegraphics[width=80mm, height=70mm] {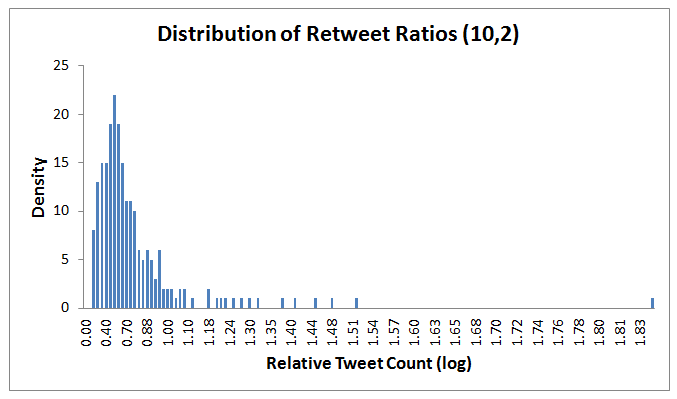}
\caption{The distribution of retweet ratios for time frame (10, 2) after the removal of tweets associated with suspected spam accounts } \label{correct}
\end{figure}

To see the difference after the posts associated with suspected spam accounts were removed, we re-calculated the distribution of user-retweet ratios again for arbitrarily chosen pairs of time frames. Figure \ref{correct} illustrates the distribution for time frames (10, 2). We observed that the distribution is now much smoother and seem to follow the log-normal distribution. We performed the log-normal test and verified that this is indeed the case. 

\subsection{Spammers and Trend-setters}

\begin{figure} [ht]
\centering
\includegraphics[width=80mm, height=70mm] {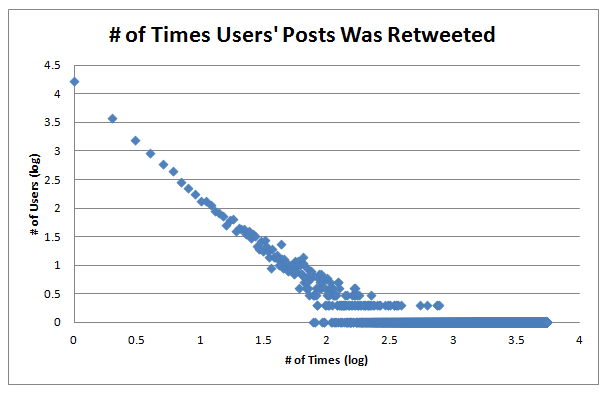}
\caption{The distribution for the frequency of retweets of user posts} \label{distribution_r}
\end{figure}

We found 6824 users in our sample whose tweets were retweeted. However, the total number of users who retweeted at least one person's tweet was 187130, which is very skewed.  Figure \ref{distribution_r} illustrates the distribution for the number of times users were retweeted. This distribution follows the power law. 

We discovered that the number of users whose tweets were retweeted by the suspected spam accounts was 4665, which is a surprising {\bf 68\%} of the users who were retweeted in our sample. This shows that the suspected spam accounts affect a majority of the trend-setters in our sample, helping them raise the retweet number of their posts and thereby making their posts appear on the trending list. The overall effect of the spammers is very significant. We also observed that a high 98\% of the total trending keywords can be found in posts retweeted by suspected spam accounts. Thus it can also be argued that {\bf many of the trends themselves are artificially generated}, which is a very important result. 

\subsection{Examples of Spam Accounts}\label{Chen}

\begin{figure} [ht]
\centering
\includegraphics[height=95mm] {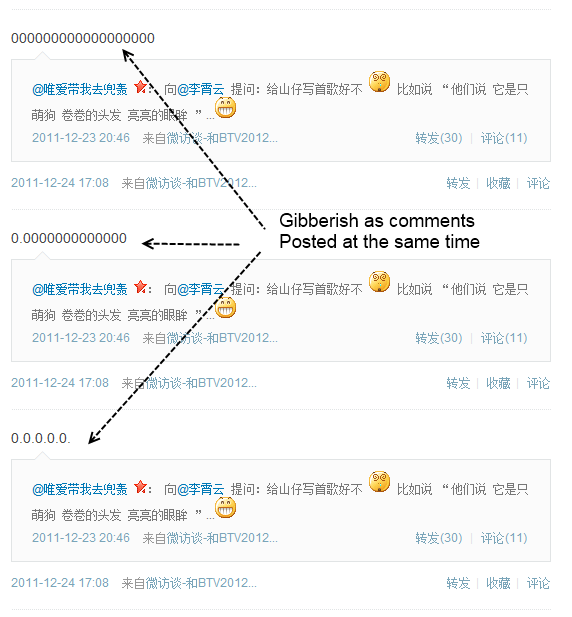}
\caption{Example of a spam account }\label{Example_2}
\end{figure}

Next, we investigate the activities of typical spam accounts in Sina Weibo. We have shown that accounts with high retweet ratios are likely to be spam accounts. Although the majority of the accounts had already been deleted by the administrator, we manually inspected 100 currently existing accounts with high retweet ratios and found that 95 clearly participate in spamming activities. The other 5 were regular users supporting their favorite singers and celebrities by repeatedly retweeting their posts, which can also be construed as spam; however, we exclude those from our list of suspected spam accounts.  Figure  \ref{Example_2} illustrates two examples of the activities from suspected spam accounts. 

First, we observe that the suspected spam accounts we inspected tend to repeatedly retweet the same post with the goal of increasing the retweet number of said post. 
Next, the interval time of these repeated retweets tend to be very close to each other with long breaks between each set. 
Finally, we observe that the replies left from spam accounts often do not make any sense (see the comments circled in Figure   \ref{Example_2}).
\cite{ChenWu} had similar findings, and explained that this was because the paid posters are mainly interested in finishing the job as quickly as possible, thus they tend to retweet multiple times in short bursts and leave gibberish as replies. We observe that the replies in \ref{Example_2} a) and b) are not proper sentences.

\begin{figure} [ht]
\centering
\includegraphics[height=90mm] {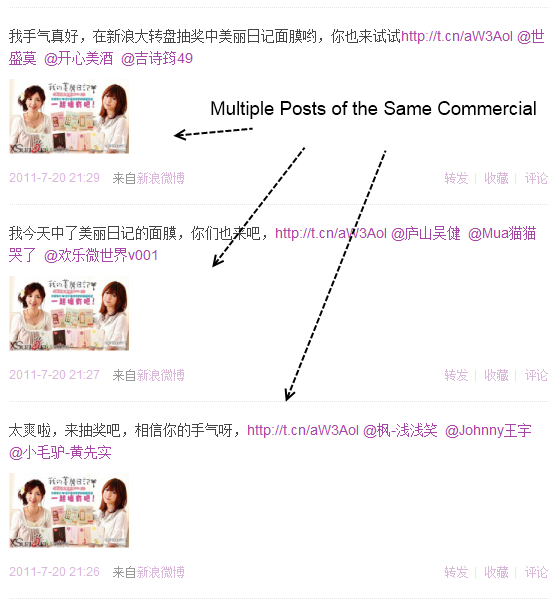}
\caption{Example of an account using spam }\label{Example_Beidong}
\end{figure}

For the 4665 users whose tweets were retweeted by at least one suspected spam account, we calculate the percentage of retweets from spam accounts and the percentage of suspected spam accounts involved. We selected only accounts whose tweets were retweeted by at least 50\% of the accounts that are suspected spam accounts. 
From our manual inspection we found mainly three types of accounts: 
\begin{enumerate}
\item Verified accounts from celebrities and reality show contestants: We hypothesize that they employ spam accounts to boast the popularity of their posts, making it seem like the posts were retweeted by many fans; 
\item Verified accounts from companies: We hypothesize that they employ spam accounts to boast the perceived popularity of their products; 
\item Unverified accounts with posts consist of ads for products: We hypothesize that these accounts employ spam accounts to distribute the ads and to boast the perceived popularity of their products, hoping other users will notice and distribute (see Figure \ref{Example_Beidong} for an example).  
\end{enumerate}

\section{Discussion and Future Work} \label{future}

We have examined the tweets relating to the trending topics in Sina Weibo. First we analyzed the growth and persistence of trends. When we looked at the distribution of tweets over time, we observed that there was a significant difference when contrasted with Twitter. The effect of retweets in Sina Weibo was significantly higher than in Twitter. 
We also found that many of the accounts that contribute to trends tend to operate as user contributed online magazines, sharing amusing pictures, jokes, stories and antidotes. Such posts tend to recieve a large amount of responses from users and thus retweets. 
Yang et al.  \cite{Yang} have shown similar results about Mitbbs users forwarding amusing messages and ``virtual gifts'' to online friends. The effect of this is similar to that of sending ``a  cyber greeting card''.  This phenomenon can also be observed from text messages sent from  cell phones between individuals in China \cite{Xia2011}.
This is interesting in the context of there being strong censorship in chinese social media. It can be hypothesized that under such circumstances, it is these kind of ``safe'' topics that can emerge. 

When we examined the retweets in more detail, we made an important discovery. We found that 49\% of the retweets in Sina Weibo containing trending keywords  were actually associated with fraudulent accounts. We observed that these accounts comprised of a small amount (1.08\% of the total users) of users but were responsible for a large percentage of the total retweets for the trending keywords. These fake accounts are responsible for artificially inflating certain posts, thus creating fake trends in Sina Weibo. 

We relate our finding to the questions we raised in the introduction. There is a strong competition among content in online social media to become popular and trend and this gives motivation to users to artificially inflate topics to gain a competitive edge. We hypothesize that certain accounts in Sina Weibo employ fake accounts to repeatedly repeat their tweets in order to propel them to the top trending list, thus gaining prominence as top trend setters (and more visible to other users). We found evidence suggesting that the accounts that do so tend to be verified accounts with commercial purposes. 

It is clear that the owners of these user contributed online magazines see this as a business opportunity to gain audience for their content. They can start by generating and propagating popular content and subsequently begin inserting advertisements amongst the jokes in their their Sina Weibo accounts. The artificial inflation makes it an even more effective campaign. 

%

We have found that we can effectively detect suspected spam accounts using retweet ratios. This can lead to future work such as using machine learning to identify other spamming techniques. In the future, we would like to examine the behavior of these fake accounts that contribute to artificial inflation in Sina Weibo to learn how successful they are in influencing trends.

\let\oldbibliography\thebibliography
\renewcommand{\thebibliography}[1]{%
  \oldbibliography{#1}%
  \setlength{\itemsep}{1 pt}%
}

\pagebreak

\begin{thebibliography}{30}
\bibitem{Asur2011}
S.~Asur, B.~A. Huberman, G.~Szabo, and C.~Wang, ``Trends in social media -
  persistence and decay,'' in \emph{5th International AAAI Conference on
  Weblogs and Social Media}, 2011.

\bibitem{Huberman}
B.~A. Huberman, D.~M. Romero, and F.~Wu, ``Social networks that matter: Twitter
  under the microscope,'' \emph{Computing Research Repository}, 2008.

\bibitem{Jin}
L.~Jin, ``{Chinese outline BBS sphere: what BBS has brought to China},''
  Master's thesis, Massachusetts Institute of Technology, April 2009.

\bibitem{flesh2}
B.~Wang, B.~Hou, Y.~Yao, and L.~Yan, ``Human flesh search model incorporating
  network expansion and gossip with feedback,'' in \emph{Proceedings of the
  2009 13th IEEE/ACM International Symposium on Distributed Simulation and Real
  Time Applications}.\hskip 1em plus 0.5em minus 0.4em\relax IEEE Computer
  Society, 2009, pp. 82--88.

\bibitem{King}
V.~King, L.~Yu, and Y.~Zhuang, ``Guanxi in the chinese web,'' in
  \emph{Proceedings of the 2009 IEEE International Conference on Computational
  Science and Engineering}, vol.~4.\hskip 1em plus 0.5em minus 0.4em\relax IEEE
  Computer Society, 2009, pp. 9--17.

\bibitem{Tai}
T.~Z. Xue, \emph{The Internet in China : Cyberspace and Civil Society}.\hskip
  1em plus 0.5em minus 0.4em\relax Routledge, 2006.

\bibitem{Statistic-general}
CNNIC. (2010) The 21st statistics report on the internet development in china
  (in chinese). [Online]. Available:
  \url{http://www.cnnic.cn/index/0E/00/11/index.htm}

\bibitem{Statistic-rural}
------. (2010) Survey report on internet development in rural china (in
  chinese). [Online]. Available:
  \url{http://www.cnnic.cn/en/index/00/02/index.htm}

\bibitem{Netizen}
F.~Y. Wang, ``Beyond x 2.0: where should we go?'' \emph{IEEE Intelligent
  Systems}, vol.~24, no.~3, pp. 2--4, 2009.

\bibitem{benevenuto2010detecting}
F.~Benevenuto, G.~Magno, T.~Rodrigues, and V.~Almeida, ``Detecting spammers on
  twitter,'' in \emph{Collaboration, Electronic messaging, Anti-Abuse and Spam
  Conference (CEAS)}, vol.~6.\hskip 1em plus 0.5em minus 0.4em\relax {National
  Academy Press}, 2010.

\bibitem{wang2010detecting}
A.~Wang, ``Detecting spam bots in online social networking sites: a machine
  learning approach,'' \emph{Data and Applications Security and Privacy XXIV},
  pp. 335--342, 2010.

\bibitem{lee2010uncovering}
K.~Lee, J.~Caverlee, and S.~Webb, ``Uncovering social spammers: social
  honeypots+ machine learning,'' in \emph{Proceeding of the 33rd international
  ACM SIGIR conference on Research and development in information
  retrieval}.\hskip 1em plus 0.5em minus 0.4em\relax {ACM}, 2010, pp. 435--442.

\bibitem{markines2009social}
B.~Markines, C.~Cattuto, and F.~Menczer, ``Social spam detection,'' in
  \emph{Proceedings of the 5th International Workshop on Adversarial
  Information Retrieval on the Web}.\hskip 1em plus 0.5em minus 0.4em\relax
  {ACM}, 2009, pp. 41--48.

\bibitem{bgh1982}
Y.~Boshmaf, I.~Muslukhov, K.~Beznosov, and M.~Ripeanu, ``The socialbot network:
  when bots socialize for fame and money,'' in \emph{Proceedings of the 27th
  Annual Computer Security Applications Conference}.\hskip 1em plus 0.5em minus
  0.4em\relax {ACM}, 2011, pp. 93--102.

\bibitem{stone2011understanding}
B.~Stone-Gross, R.~Stevens, A.~Zarras, R.~Kemmerer, C.~Kruegel, and G.~Vigna,
  ``Understanding fraudulent activities in online ad exchanges,'' in
  \emph{Proceedings of the 2011 ACM SIGCOMM conference on Internet measurement
  conference}.\hskip 1em plus 0.5em minus 0.4em\relax {ACM}, 2011, pp.
  279--294.

\bibitem{yu2010sbotminer}
F.~Yu, Y.~Xie, and Q.~Ke, ``Sbotminer: large scale search bot detection,'' in
  \emph{Proceedings of the third ACM international conference on Web search and
  data mining}.\hskip 1em plus 0.5em minus 0.4em\relax {ACM}, 2010, pp.
  421--430.

\bibitem{6406086}
L.~Liu and K.~Jia, ``Detecting spam in chinese microblogs - a study on sina
  weibo,'' in \emph{Computational Intelligence and Security (CIS), 2012 Eighth
  International Conference on}, 2012, pp. 578--581.

\bibitem{6425674}
Y.~Zhou, K.~Chen, L.~Song, X.~Yang, and J.~He, ``Feature analysis of spammers
  in social networks with active honeypots: A case study of chinese
  microblogging networks,'' in \emph{Advances in Social Networks Analysis and
  Mining (ASONAM), 2012 IEEE/ACM International Conference on}, 2012, pp.
  728--729.

\bibitem{XuChen}
X.~Yong, Z.~Yi, and C.~Kai, ``Observation on spammers in sina weibo,'' in
  \emph{Proceedings of the 2nd International Conference on Computer Science and
  Electronics Engineering (ICCSEE 2013)}.\hskip 1em plus 0.5em minus
  0.4em\relax Atlantis Press, 2013.

\bibitem{Lin:2013:AIS:2501025.2501035}
C.~Lin, J.~He, Y.~Zhou, X.~Yang, K.~Chen, and L.~Song, ``Analysis and
  identification of spamming behaviors in sina weibo microblog,'' in
  \emph{Proceedings of the 7th Workshop on Social Network Mining and Analysis},
  ser. SNAKDD '13, 2013, pp. 5:1--5:9.

\bibitem{ChenWu}
C.~Chen, K.~Wu, V.~Srinivasan, and X.~Zhang, ``Battling the internet water
  army: Detection of hidden paid posters,'' \emph{CoRR}, vol. abs/1111.4297,
  2011.

\bibitem{Jamali}
M.~Jamali and H.~Abolhassani, ``Different aspects of social network analysis,''
  in \emph{Proceedings of the 2006 IEEE/WIC/ACM International Conference on Web
  Intelligence}, 2006, pp. 66--72.

\bibitem{Mislove}
A.~Mislove, M.~Marcon, K.~P. Gummadi, P.~Druschel, and B.~Bhattacharjee,
  ``Measurement and analysis of online social networks,'' in \emph{Proceedings
  of the 7th SIGCOMM Conference on Internet Measurement}.\hskip 1em plus 0.5em
  minus 0.4em\relax ACM, 2007, pp. 29--42.

\bibitem{Buchanan}
M.~Buchanan, \emph{Nexus: Small Worlds and the Groundbreaking Theory of
  Networks}.\hskip 1em plus 0.5em minus 0.4em\relax {W. W. Norton \& Company},
  May 2003.

\bibitem{Kumar}
R.~Kumar, J.~Novak, and A.~Tomkins, ``Structure and evolution of online social
  networks,'' in \emph{Proceedings of the 12th ACM SIGKDD International
  Conference on Knowledge Discovery and Data Mining}.\hskip 1em plus 0.5em
  minus 0.4em\relax ACM, 2006, pp. 611--617.

\bibitem{mcpherson2001birds}
M.~McPherson, L.~Smith-Lovin, and J.~M. Cook, ``Birds of a feather: homophily
  in social networks,'' \emph{Annual Review of Sociology}, vol.~27, no.~1, pp.
  415--444, 2001.

\bibitem{Agarwal2008Identifying}
N.~Agarwal, H.~Liu, L.~Tang, and P.~S. Yu, ``{Identifying the Influential
  Bloggers in a Community},'' \emph{WSDM'08}, 2008.

\bibitem{Backstrom}
L.~Backstrom, D.~Huttenlocher, J.~Kleinberg, and X.~Lan, ``Group formation in
  large social networks: membership, growth, and evolution,'' in
  \emph{Proceedings of the 12th International Conference on Knowledge Discovery
  and Data Mining}.\hskip 1em plus 0.5em minus 0.4em\relax ACM, 2006, pp.
  44--54.

\bibitem{Crandall}
D.~Crandall, D.~Cosley, D.~Huttenlocher, J.~Kleinberg, and S.~Suri, ``Feedback
  effects between similarity and social influence in online communities,'' in
  \emph{Proceedings of the 14th ACM SIGKDD international conference on
  Knowledge discovery and data mining}.\hskip 1em plus 0.5em minus 0.4em\relax
  ACM, 2008, pp. 160--168.

\bibitem{Romero2011}
D.~M. Romero, W.~Galuba, S.~Asur, and B.~A. Huberman, ``Influence and passivity
  in social media,'' in \emph{20th International World Wide Web Conference
  (WWW'11)}, 2011.

\bibitem{Kwak}
H.~Kwak, C.~Lee, H.~Park, and S.~Moon, ``What is twitter, a social network or a
  news media?'' in \emph{Proceedings of the 19th international conference on
  World wide web}, ser. WWW '10, 2010, pp. 591--600.

\bibitem{Mathioudakis}
M.~Mathioudakis and N.~Koudas, ``Twittermonitor: trend detection over the
  twitter stream,'' in \emph{Proceedings of the 2010 international conference
  on Management of data}, ser. SIGMOD '10, 2010, pp. 1155--1158.

\bibitem{Wu2}
S.~Wu, J.~M. Hofman, W.~A. Mason, and D.~J. Watts, ``Who says what to whom on
  twitter,'' in \emph{Proceedings of the 20th international conference on World
  wide web}, ser. WWW '11, 2011, pp. 705--714.

\bibitem{asur2011trends}
S.~Asur, B.~A. Huberman, G.~Szabo, and C.~Wang, ``Trends in social media:
  Persistence and decay,'' in \emph{5th International AAAI Conference on
  Weblogs and Social Media}.\hskip 1em plus 0.5em minus 0.4em\relax {AAAI},
  2011, pp. 434--437.

\bibitem{cha2010measuring}
M.~Cha, H.~Haddadi, F.~Benevenuto, and K.~P. Gummadi, ``Measuring user
  influence in twitter: The million follower fallacy,'' in \emph{4th
  International AAAI Conference on Weblogs and Social Media (ICWSM)}.\hskip 1em
  plus 0.5em minus 0.4em\relax {AAAI}, 2010.

\bibitem{StrongTie}
Y.~Bian, ``Bringing strong ties back in: indirect ties, network bridges, and
  job searches in china,'' \emph{American Sociological Review}, vol.~62, no.~3,
  pp. 366--385, 1997.

\bibitem{WorkControl}
D.~Ruan, ``Interpersonal networks and workplace controls in urban china,''
  \emph{The Australian Journal of Chinese Affairs}, vol.~29, pp. 89--105, 1993.

\bibitem{768262}
J.-L. Farh, A.~S. Tsui, K.~Xin, and B.-S. Cheng, ``The influence of relational
  demography and guanxi: the {C}hinese case,'' \emph{Organization Science},
  vol.~9, no.~4, pp. 471--488, 1998.

\bibitem{Guanxi3}
Y.~Bian, R.~Breiger, D.~Davis, and J.~Galaskiewicz, ``Occupation, class, and
  social networks in urban china,'' \emph{Social Forces}, vol.~83, no.~4, pp.
  1443--1468, 2005.

\bibitem{Carrington}
P.~J. Carrington, J.~Scott, and S.~Wasserman, Eds., \emph{Models and Methods in
  Social Network Analysis}.\hskip 1em plus 0.5em minus 0.4em\relax Cambrige
  University Press, 2005.

\bibitem{Xin}
M.~Xin, ``Chinese bulletin board system's influence upon university students
  and ways to cope with it (in chinese),'' \emph{Journal of Nanjing University
  of Technology (Social Science Edition)}, vol.~4, pp. 100 --104, 2003.

\bibitem{yu}
L.~Yu and V.~King, ``The evolution of friendships in chinese online social
  networks,'' in \emph{Proceedings of the 2010 IEEE Second International
  Conference on Social Computing}, ser. SOCIALCOM '10, 2010, pp. 81--87.

\bibitem{Zhong2010}
Z.-J. Zhong, ``Social networking services (sns) in china,'' \emph{International
  Journal of e-Business Management}, vol.~4, no.~1, pp. 66--69, 2010.

\bibitem{Zhang201215}
B.~Zhang, X.~Guan, M.~J. Khan, and Y.~Zhou, ``A time-varying propagation model
  of hot topic on \{BBS\} sites and blog networks,'' \emph{Information
  Sciences}, vol. 187, no.~0, pp. 15 -- 32, 2012.

\bibitem{Chan2012}
M.~Chan, X.~Wu, Y.~Hao, R.~Xi, and T.~Jin, ``Microblogging, online expression,
  and political efficacy among young chinese citizens: The moderating role of
  information and entertainment needs in the use of weibo,'' \emph{Cyberpsy.,
  Behavior, and Soc. Networking}, vol.~15, no.~7, pp. 345--349, 2012.

\bibitem{Chen2012}
H.~Chen and E.~Haley, ``The lived meanings of product placement in social
  network sites (snss) among urban chinese white-collar professional users: A
  story of happy network,'' \emph{Journal of Interactive Advertising}, vol.~11,
  no.~1, pp. 11--16, 2010.

\bibitem{Chu2011}
S.-C. Chu and S.~M. Choi, ``Electronic word-of-mouth in social networking
  sites: A cross-cultural study of the united states and china,'' \emph{Journal
  of Global Marketing}, vol.~24, no.~3, pp. 263--281, 2011.

\bibitem{Yang}
J.~Yang, M.~S. Ackerman, and L.~A. Adamic, ``Virtual gifts and guanxi:
  Supporting social exchange in a chinese online community,'' in
  \emph{Proceedings of the ACM 2011 Conference on Computer Supported
  Cooperative Work}, ser. CSCW '11, 2011, pp. 45--54.

\bibitem{Lin}
J.~Lin, Z.~Li, D.~Wang, K.~Salamatian, and G.~Xie, ``Analysis and comparison of
  interaction patterns in online social network and social media,'' in
  \emph{Computer Communications and Networks (ICCCN), 2012 21st International
  Conference on}, 2012, pp. 1--7.

\bibitem{Renren}
J.~Niu, J.~Peng, L.~Shu, C.~Tong, and W.~Liao, ``An empirical study of a
  chinese online social network--renren,'' \emph{Computer}, vol.~46, no.~9, pp.
  78--84, 2013.

\bibitem{Chang2}
G.~Chang, H.-S. Huang, and J.~Y. jen Hsu, ``Detecting chinese wish messages in
  social media: An empirical study,'' in \emph{ICWSM}, 2013.

\bibitem{Fan}
R.~Fan, J.~Zhao, Y.~Chen, and K.~Xu, ``Anger is more influential than joy:
  Sentiment correlation in weibo,'' \emph{CoRR}, vol. abs/1309.2402, 2013.

\bibitem{Xia2011}
Y.~Xia, ``Chinese use of mobile texting for social interactions: Cultural
  implications in the use of communication technology,'' \emph{Intercultural
  Communication Studies}, vol.~21, no.~2, p. 131, 2012.

\end{thebibliography}
\end{document}